\documentclass[10pt, prd,aps,twocolumn,%
showpacs,preprintnumbers,amsmath,amssymb]{revtex4-2}
\usepackage[utf8]{inputenc}
\usepackage{graphicx}
\usepackage{bm}
\usepackage{siunitx}
\usepackage[breaklinks, colorlinks=true]{hyperref}

\newcommand{\kel}{k_2}
\newcommand{\kmag}{j_2}

\begin{document}

\title{
Magnetic-type Love number differentiating quark stars from neutron stars
}

\author{Kenji Fukushima}
\email{fuku@nt.phys.s.u-tokyo.ac.jp}
\affiliation{Department of Physics, The University of Tokyo, 
  7-3-1 Hongo, Bunkyo-ku, Tokyo 113-0033, Japan}

\author{Josuke Minamiguchi}
\email{minamiguchi@nt.phys.s.u-tokyo.ac.jp}
\affiliation{Department of Physics, The University of Tokyo, 
  7-3-1 Hongo, Bunkyo-ku, Tokyo 113-0033, Japan}

\author{Tomoya Uji}
\email{uji@nt.phys.s.u-tokyo.ac.jp}
\affiliation{Department of Physics, The University of Tokyo, 
  7-3-1 Hongo, Bunkyo-ku, Tokyo 113-0033, Japan}

\begin{abstract}
  The quark star (QS) is a hypothetical and yet undiscovered stellar object, and its existence would mark a paradigm shift in research on nuclear and quark matter.  Although compactness is a well-known signature for distinguishing between two branches of QSs and neutron stars (NSs), some QSs can overlap with NSs in the radius-mass plane.  To manifest their evident differences, we investigate the tidal properties of QSs and NSs.  We then find that the magnetic-type Love number is a robust indicator for differentiating between QSs and NSs, whereas the electric-type one is insufficient when QSs and NSs have similar masses and radii.
  Finally, we show that gravitational waves from binary star mergers can, in principle, be sensitive to differences between QSs and NSs.
\end{abstract}

\maketitle

\paragraph*{Introduction:}
Strange quark matter (SQM) could be more stable than hadronic matter at zero pressure~\cite{Bodmer:1971we, Farhi:1984qu, Witten:1984rs}.
According to this SQM hypothesis, compact stellar objects composed entirely of quark matter, i.e., quark stars (QSs), can exist in the universe~\cite{Itoh:1970uw, Fechner:1978ji, Alcock:1986hz, Haensel:1986qb, Bombaci:2004mt, Jaikumar:2005ne, Weber:2004kj}.
Numerous attempts are made 
(see Refs.\!~\cite{Sotani:2003zc,Fu:2008bu} for $f$- and $g$-mode oscillations,
Ref.\!~\cite{Watts:2006hk} for a critical discussion on seismic vibrations, Ref.~\cite{Traversi:2020dho} for machine learning, Ref.~\cite{Zou:2022voz} for core-crust oscillation, etc.) and more exotic scenarios are discussed~\cite{Drago:2013fsa, Li:2016khf, Holdom:2017gdc, Bombaci:2020vgw, Zhang:2020jmb}, but no convincing support for the QS existence has been established.
Recently, QS studies have widely attracted interest since the compact object in the supernova remnant, HESS J1731-347~\cite{Kosack:2007zz}, may have $M=0.77^{+0.20}_{-0.17}M_\odot$ and $R=10.4^{+0.86}_{-0.78}\,\mathrm{km}$~\cite{Doroshenko:2022nwp}, where $M$ and $R$ are the mass and the radius of the star and $M_\odot$ is the solar mass, implying that it is either the lightest neutron star (NS) or it may be the QS candidate~\cite{Klochkov:2014ola}. 

Because first-principles QCD at high density remains inaccessible owing to the sign problem (see Ref.~\cite{Nagata:2021ugx,Fujimoto:2024cyv} for a review), compact-star EoSs must be constrained experimentally.
Thanks to advances in both perturbative QCD at high density~\cite{Kurkela:2009gj,Gorda:2018gpy} and astrophysical observation analyses~\cite{Fujimoto:2021zas,Altiparmak:2022bke,Marczenko:2022jhl,Brandes:2023hma}, the distribution of masses and radii of the NSs can constrain the EoS well up to $\sim 5$ times the nuclear saturation density.  In this way, it is likely that a core of quark matter exists in massive NSs~\cite{Annala:2019puf}, which can be further tested by the future gravitational-wave (GW) measurements~\cite{Most:2018eaw,Weih:2019xvw,Bauswein:2022vtq,Fujimoto:2022xhv,Fujimoto:2024ymt}.  Furthermore, GW170817 supposedly originates from the merger of binary NSs, and the inferred value of the dimensionless tidal deformability, $\Lambda$, provides additional constraints on the likely EoS candidate~\cite{Annala:2017llu,LIGOScientific:2017vwq}.

In the near future, more and more GW signals will be detected and a useful set of $\{M,\, \kel\}$ will be accumulated, where $\kel$ is the electric-type quadrupolar Love number leading to $\Lambda$~\cite{Hinderer:2007mb}:
\begin{align}
    k_2 = \frac{3}{2}\Lambda\left(\frac{M}{R}\right)^{\!5} \,,
    \qquad
    \Lambda = \frac{\lambda}{M^5} \,,
\end{align}
where $c = G_N = 1$, and the dimensionless deformability is $\Lambda = G_N\lambda/(G_N M)^5$ if $G_N$ is restored.  Here, $\lambda$ is the induced-mass-quadrupole response coefficient.
This can be translated into a set of $\{M,\, R\}$.  
In the optimistic scenario, some $M$ and/or $R$ far apart from the $M$-$R$ branch of the NSs would be a clear signature for the QS\@.  
However, the QSs may have a distribution in the larger mass region
and their preferred mass may be around $\gtrsim 1.4M_\odot$ like the NSs or even heavier as discussed in the GW190814 context~\cite{Zhou:2017pha, Traversi:2020dho, Cao:2020zxi, Li:2020wbw, Miao:2021nuq, Miao:2021gmf, Miao:2024qik, Shirke:2025gfi}.  
In other words, some objects that are considered to be massive NSs could be the QSs masquerading as NSs (see Ref.~\cite{Alford:2004pf} for related discussions on a hybrid star).  
If we find any indicator to distinguish between the QSs and the NSs in the mass region $\gtrsim 1.4M_\odot$, it would crucially enhance our chance to discover the QS candidates.

To disentangle similarity between massive QSs and NSs, we need to introduce another observable in addition to $\{M,\, R\}$ or $\{M,\, \kel\}$.  An immediate candidate is the combination of all, i.e., $\{M,\, R,\, \kel\}$.  
However, this idea does not work well.
To obtain $\{M,\, R,\, \kel\}$, it is necessary to determine each of them independently, which requires simultaneous observations of electromagnetic and GWs from the same star.  There is no such simultaneous observation, and the prospect of future observation will be extremely limited.

One may think that the Love-$C$ relation~\cite{Yagi:2016bkt} complements $\{M,\, R,\, \kel\}$, but this approach does not work, either.  
The conventional Love-$C$ relation is derived from various NS EoSs, and thus, only the NS combinations of $\{M,\, R,\, \kel\}$ should inevitably be concluded as a result of the Love-$C$ relation applied to $\{M,\, R\}$ or $\{M,\, \kel\}$.  To overcome this, one may find a broader Love-$C$ relation that can fit the EoSs of both QSs and NSs.  In that case, however, such a broader Love-$C$ relation is no longer precise, and the uncertainty diminishes the resolution to distinguish between the QSs and the NSs from $\{M,\, R,\, \kel\}$.

In our work, we propose that the magnetic-type quadrupolar Love number, $\kmag$, is a new distinguishing indicator.  We note that $\kmag$ carries information independent of $\kel$ imprinting on the GW signals, and in our convention, the definition of $\kmag$ follows from Ref.~\cite{Damour:2009vw}, i.e.,
\begin{align}
    j_2 = 48\,\Sigma\left(\frac{M}{R}\right)^{\!5} \,,
    \qquad
    \Sigma \equiv \frac{\sigma}{M^5} \,,
\end{align}
where the dimensionless magnetic-type tidal deformability is $\Sigma = G_N\sigma/(G_N M)^5$ if $G_N$ is restored.  Here, $\sigma$ is the
induced-current-quadrupole response coefficient.
The magnetic-type Love number, $\kmag$, is estimated from the given EoS, and it characterizes a response of a self-gravitating body to external magnetic tidal fields corresponding to parity-odd perturbations~\cite{Damour:2009vw, Landry:2015cva, Binnington:2009bb}. 

We shall demonstrate that $\kmag$ is a promising indicator for a wide variety of SQM EoSs.  In the literature~\cite{Yagi:2016bkt,Yagi:2013sva, Postnikov:2010yn}, some SQM EoSs were adopted from phenomenological models for the QSs, and we extensively generalize the analysis using 28 independent QS EoSs.
For comparison, we employ 28 different NS EoSs.  
In this way, we estimate the possible ranges of $\kel$ and $\kmag$ belonging to the QS and NS branches, respectively.
Our results show that, even when the QS and NS branches stay close in the $M$-$R$ plane or the $M$-$\kel$ plane, two branches are well separated in the $M$-$\kmag$ plane.
\vspace{0.5em}

\begin{figure}
    \centering
    \includegraphics[width=0.98\linewidth]{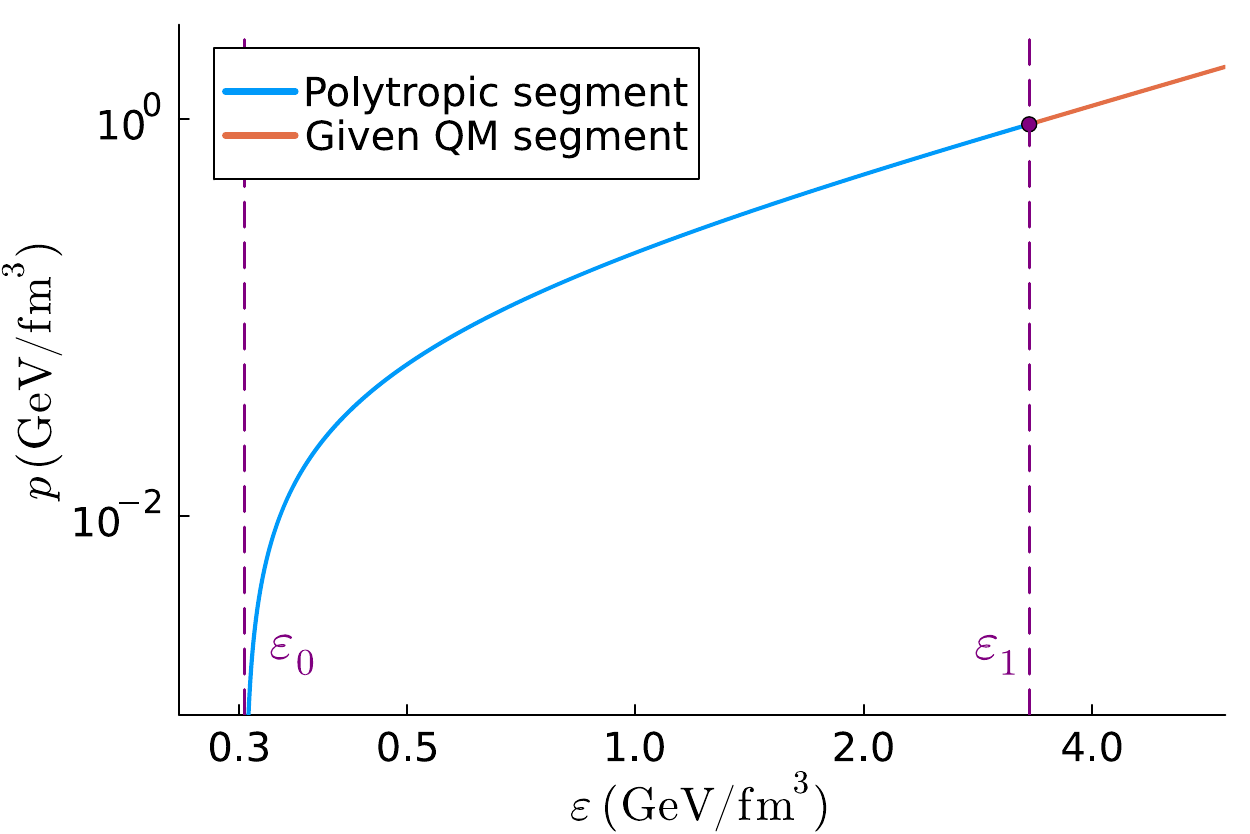}
\caption{Parametrization of the QS EoS with $\varepsilon_0$ and $\varepsilon_1$.
The orange curve represents the high density part, $p_\mathrm{high}(\varepsilon)$.
For a given pair of $\varepsilon_0$ and $\varepsilon_1$,
the polytropic part (blue curve) is adjusted so that the QS EoS is smooth at $\varepsilon=\varepsilon_1$.}
\label{fig:para_EoS}
\end{figure}

\paragraph{Construction of SQM EoSs:}
Our working hypothesis is that QSs consist entirely of SQM~\cite{Alcock:1986hz,Haensel:1986qb, Bombaci:2004mt, Jaikumar:2005ne}.
Since the surface of QSs, defined by $p=0$, is composed of SQM, such EoSs have a nonzero surface energy density; $\varepsilon_0 \neq 0$.
The value of $\varepsilon_0$ should be large enough to stabilize SQM as the true ground state.
We construct a broad range of SQM EoS candidates in a way that satisfies physical requirements.
Specifically, we adopt the following prescriptions to generate SQM EoS candidates:

\begin{itemize}
  \item \textbf{Asymptotic Boundary Condition $(\varepsilon \ge \varepsilon_1)$}:
        For high enough energy densities above a threshold, $\varepsilon_1$, the pressure is assumed to be given by the pQCD or NJL model calculation, i.e.,
        $p = p_{\mathrm{high}}(\varepsilon)$.
  \item \textbf{Polytropic Interpolation $(\varepsilon_0\!\le \varepsilon\!\le \varepsilon_1)$}:
        Between $\varepsilon_0$ at the stellar surface and the threshold $\varepsilon_1$, a single polytropic parametrization is assumed as
        $p = K\bigl(\varepsilon-\varepsilon_{0}\bigr)^{\!\alpha}$.
\end{itemize}
There are four parameters, $K$, $\alpha$, $\varepsilon_0$, and $\varepsilon_1$, and they are constrained by the smoothness condition and the causality condition.
The Fig.~\ref{fig:para_EoS} shows an example in which pQCD is chosen as the EoS on the high-density side.

The smoothness condition means that neither $p$ nor $dp/d\varepsilon$ has a discontinuity at $\varepsilon=\varepsilon_1$.  
That is,
\begin{align}
    K(\varepsilon-\varepsilon_0)^{\alpha} \Bigr|_{\varepsilon=\varepsilon_1} &= p_{\mathrm{high}}(\varepsilon) \Bigr|_{\varepsilon=\varepsilon_1} \,,
    \label{eq:cont_cond1}\\
    \frac{d}{d\varepsilon} K(\varepsilon-\varepsilon_0)^{\alpha} \Bigr|_{\varepsilon=\varepsilon_1} &= \frac{d p_{\mathrm{high}}(\varepsilon)}{d\varepsilon} \Bigr|_{\varepsilon=\varepsilon_1} \,.
    \label{eq:cont_cond2}
\end{align}
These conditions fix $K$ and $\alpha$ for given $\varepsilon_0$ and $\varepsilon_1$.
        
The causality condition reads that the speed of sound, $c_s^{2} = d p/d\varepsilon$, never exceeds the speed of light, i.e., unity in the present units, for any energy density.  Since $p=p_{\mathrm{high}}(\varepsilon \ge \varepsilon_1)$ is consistent with causality, this condition restricts a physical range of $\varepsilon_0$ and $\varepsilon_1$.  For example, if the window between $\varepsilon_0$ and $\varepsilon_1$ is too narrow, the causality condition would be violated.

To generate concrete SQM EoSs based on the aforementioned construction, we need to fix $\varepsilon_0$ and $\varepsilon_1$.  The allowed range of $\varepsilon_0$ is determined as follows.
According to the MIT bag model, $\varepsilon_0$ is related to the bag pressure, $B$, as $\varepsilon_0 = 4B$~\cite{Chodos:1974je}.
Then, $B$ must have a lower bound to prevent atomic nuclei from decaying into non-strange QM\@.  The minimum value of $B$ is estimated to be $B=B_\mathrm{min}\simeq \SI{57.5}{MeV/fm^3}$~\cite{Farhi:1984qu, DeGrand:1975cf}, leading to $\varepsilon_0 \gtrsim 4B_\mathrm{min} \simeq \SI{230}{MeV/fm^3}$.  
It is more difficult to constrain the upper bound on $\varepsilon_0$, but we do not need it in the present context.  If $\varepsilon_0$ is excessively large, the gap between QSs and NSs in the $M$-$R$ plane would become wider.  Then, it would be easier to differentiate them even without referring to the tidal properties.  
In addition to $\varepsilon_0$, we must set $\varepsilon_1$ in such a way as not to violate the causality condition.
Because 
$dp/d\varepsilon = K\alpha(\varepsilon-\varepsilon_0)^{\alpha-1}$, it is clear that $\alpha > 1$ is a necessary condition; otherwise, $c_s$ diverges at $\varepsilon = \varepsilon_0$.

\begin{table}
    \centering
    \begin{tabular}{ccc}
    \hline\hline
     Quark Matter EoS~ 
     & ~$\varepsilon_0\ (\mathrm{GeV/fm^3})$~ 
     & ~$\varepsilon_1-\varepsilon_0\ (\mathrm{GeV/fm^3})$\\
     \hline
      NJL $(g_V^{(0)},H^{(0)})$ & $0.23$ & $1$ \\
      NJL $(g_V^{(0)},H^{(0)})$ & $0.23$ & $2$ \\
      NJL $(g_V^{(0)},H^{(0)})$ & $0.23$ & $3$ \\
      NJL $(g_V^{(0)},H^{(0)})$ & $0.3052$ & $1$ \\
      NJL $(g_V^{(0)},H^{(0)})$ & $0.3052$ & $2$ \\
      NJL $(g_V^{(0)},H^{(0)})$ & $0.3052$ & $3$ \\
      NJL $(g_V^{(1)},H^{(1)})$ & $0.23$ & $1$ \\
      NJL $(g_V^{(1)},H^{(1)})$ & $0.23$ & $2$ \\
      NJL $(g_V^{(1)},H^{(1)})$ & $0.23$ & $3$ \\
      NJL $(g_V^{(1)},H^{(1)})$ & $0.3052$ & $1$ \\
      NJL $(g_V^{(1)},H^{(1)})$ & $0.3052$ & $2$ \\
      NJL $(g_V^{(1)},H^{(1)})$ & $0.3052$ & $3$ \\
      NJL $(g_V^{(2)},H^{(2)})$ & $0.23$ & $1$ \\
      NJL $(g_V^{(2)},H^{(2)})$ & $0.23$ & $2$ \\
      NJL $(g_V^{(2)},H^{(2)})$ & $0.23$ & $3$ \\
      NJL $(g_V^{(2)},H^{(2)})$ & $0.3052$ & $1$ \\
      NJL $(g_V^{(2)},H^{(2)})$ & $0.3052$ & $2$ \\
      NJL $(g_V^{(2)},H^{(2)})$ & $0.3052$ & $3$ \\
      pQCD & $0.23$ & $2$ \\
      pQCD & $0.23$ & $3$ \\
      pQCD & $0.23$ & $4$ \\
      pQCD & $0.23$ & $5$ \\
      pQCD & $0.23$ & $\infty$ \\
      pQCD & $0.3052$ & $3$ \\
      pQCD & $0.3052$ & $4$ \\
      pQCD & $0.3052$ & $5$\\
      pQCD & $0.3052$ & $\infty$ \\
      \hline\hline
    \end{tabular}
    \caption{Parameters for the SQM EoSs.  The quark matter part is given by the NJL model in Ref.~\cite{Baym:2017whm} with several coupling choices,
    $g_V^{(0)}/G=0.5$, $g_V^{(1)}/G=0.8$, $g_V^{(2)}/G=1.0$,
    $H^{(0)}/G=1.4$, $H^{(1)}/G=1.5$, $H^{(2)}/G=1.6$, where $G$ represents the scalar-pseudoscalar coupling.
    Another choice of the quark matter part is the pQCD-based EoS with $X=4$ using the notation in Ref.~\cite{Fraga:2013qra}.
    For the choice of $\varepsilon_0=4B$, $\SI{0.23}{GeV/fm^3}$ is taken from Refs.~\cite{Farhi:1984qu, DeGrand:1975cf} and $\SI{0.3052}{GeV/fm^3}$ is taken from Ref.~\cite{Buballa:2003qv}.
    For the choice of $\varepsilon_1\to\infty$, practically, $\varepsilon_1-\varepsilon_0=10^3\,\mathrm{GeV/fm^3}$ is chosen. 
    } 
    \label{tab:params}
\end{table}

More specifically, for QSs, we use three Nambu--Jona-Lasinio (NJL) models for $p_\mathrm{high}(\varepsilon)$, with model parameters taken from Ref.~\cite{Baym:2017whm}.
For each choice of the vector interaction $g_V$ and the diquark interaction $H$, we change $\varepsilon_0$ and $\varepsilon_1$ as listed in Table~\ref{tab:params}.
Also, a pQCD-based EoS is employed with $X=4$, where $X$ is the scale parameter in Ref.~\cite{Fraga:2013qra}.
For recent higher-order calculations of the cold-quark-matter pressure, see also Refs.~\cite{Gorda:2023mkk,Navarrete:2024zgz,Fernandez:2024ilg,Karkkainen:2025nkz}.
Furthermore, for reference in later discussions, we adopt the MIT bag model with $B = \SI{57.5}{MeV/fm^3}$ for the QS EoS~\cite{Farhi:1984qu, DeGrand:1975cf}.

The construction of the NS EoS sample is straightforward.
We take typical NS EoSs from the polytropic parametrization listed in  Ref.~\cite{Read:2008iy}. 
There are 34 EoSs in the original listing, but we exclude APR1--4, ENG, WFF1--2, BBB2, and GS1a, which violate the causality condition within the energy region up to eight times the energy density at nuclear saturation.
The NS EoSs used here are drawn from a broader catalog that includes not only purely nucleonic matter but also EoSs with hyperons, pion or kaon condensates, and quark components, including CFL quark matter.
As a robustness check for a strong first-order phase transition, we also overlay the hybrid-star EoSs BBKF RDF 1.2 and 1.9~\cite{Hempel:2009mc,Bastian:2020unt}, downloaded from the CompOSE database~\cite{Typel:2013rza,Oertel:2016bki,CompOSECoreTeam:2022ddl}.
As a guide to the eyes, we also plot the results from the Crossover EoS taken from Ref.~\cite{Fujimoto:2022xhv}.
\vspace{0.5em}

\paragraph{Indistinguishability of massive QSs and NSs:}

\begin{figure}
    \centering
    \includegraphics[width=0.98 \linewidth]{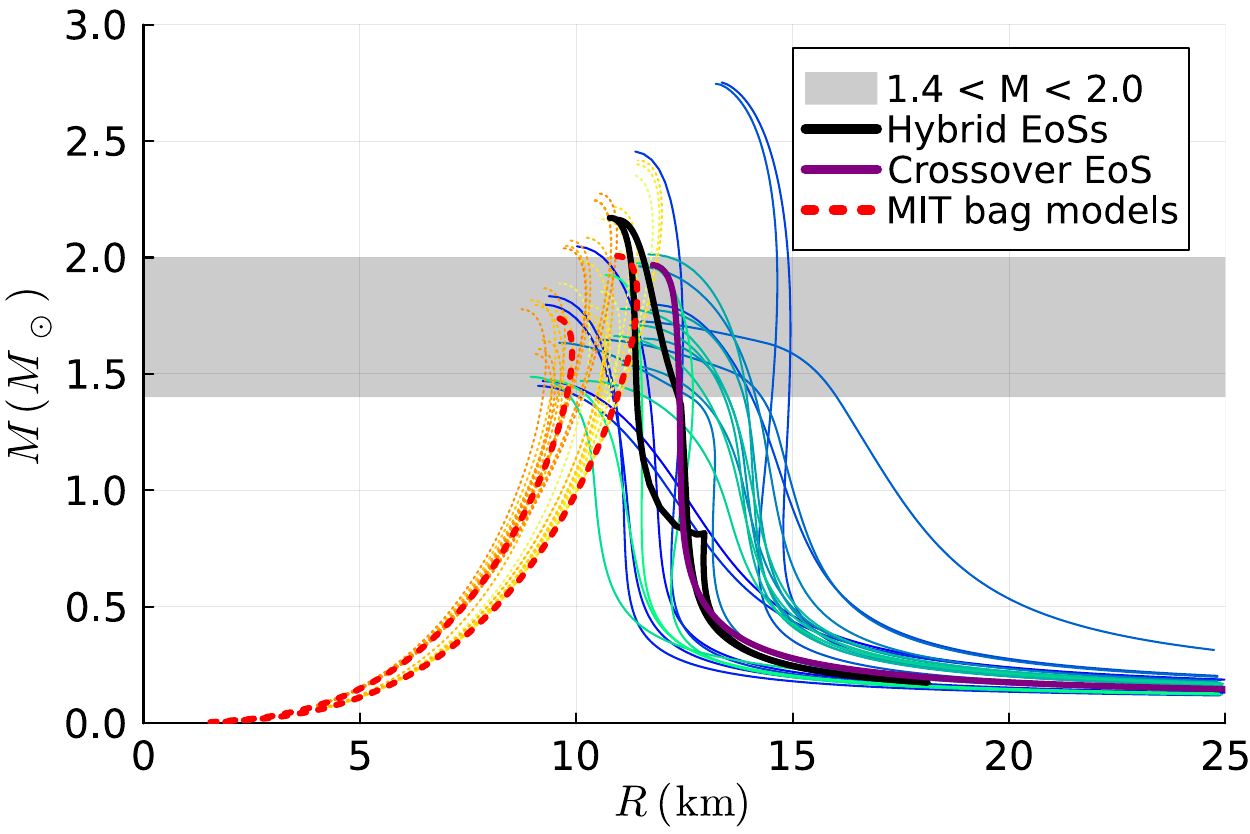}
    \caption{$M$-$R$ curves for NSs (solid) and QSs (dotted). 
    The gray band indicates the mass region favored by current NS observations. 
    }
    \label{fig:mr}
\end{figure}

\begin{figure}
    \centering
    \includegraphics[width=0.98 \linewidth]{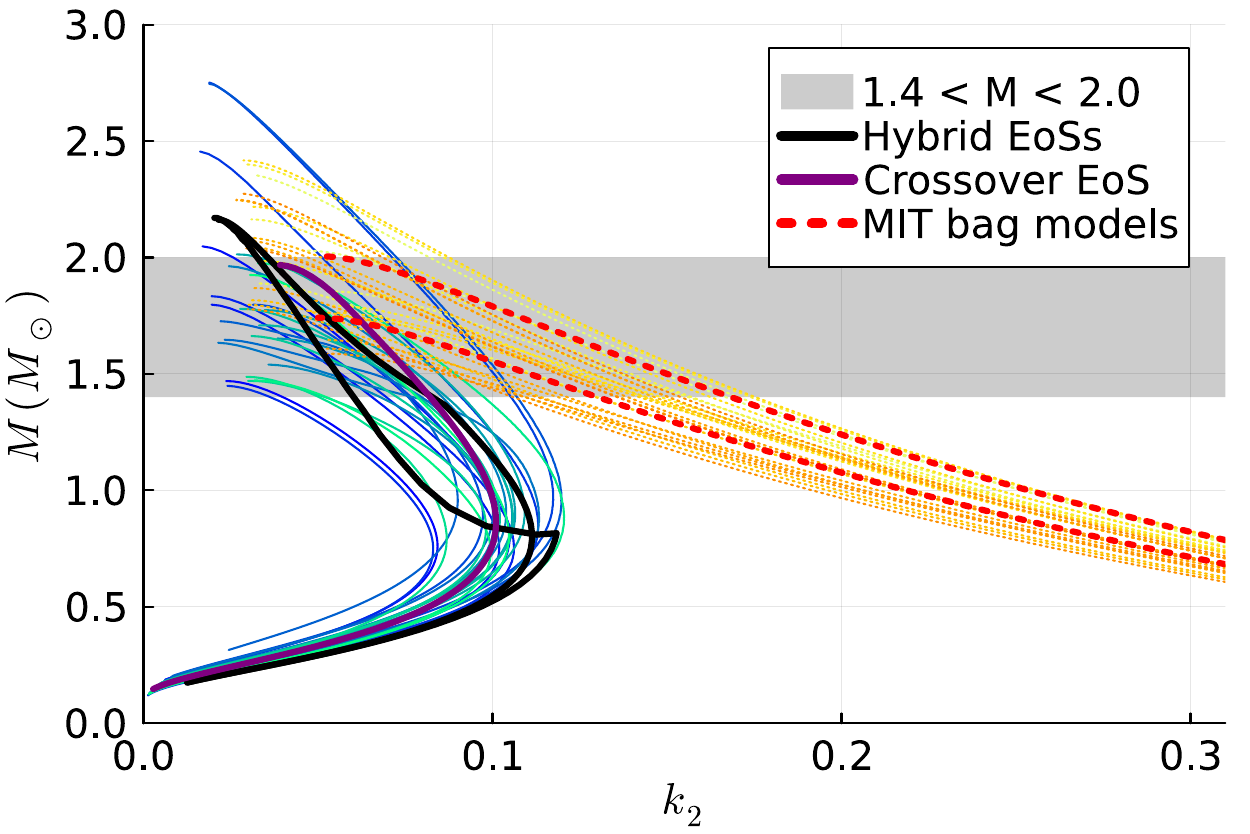}
    \caption{$M$-$\kel$ curves for NSs (solid) and QSs (dotted). 
    }
    \label{fig:mkel}
\end{figure}

Figs~\ref{fig:mr} and \ref{fig:mkel} show that QS and NS branches overlap in both the $M$-$R$ and $M$-$\kel$ planes.
It is impossible to distinguish QSs from NSs solely based on these observations in the mass region $M \gtrsim 1.4\,M_\odot$.

Our results that $\kel$ takes similar values for QSs and NSs in the high-mass range are consistent with the discussions in Ref.~\cite{Postnikov:2010yn}.
\vspace{0.5em}

\paragraph{Magnetic-type Love number:}
Following previous studies~\cite{Damour:2009vw, Landry:2015cva, Favata:2005da, Shapiro:1996up}, we assume an irrotational fluid.
Even if we calculate $\kmag$ under the constraint of strict hydrostatic equilibrium following Refs.~\cite{Binnington:2009bb}, our conclusion about $\kmag$ is qualitatively unchanged. 

\begin{figure}
    \centering
    \includegraphics[width=0.98 \linewidth]{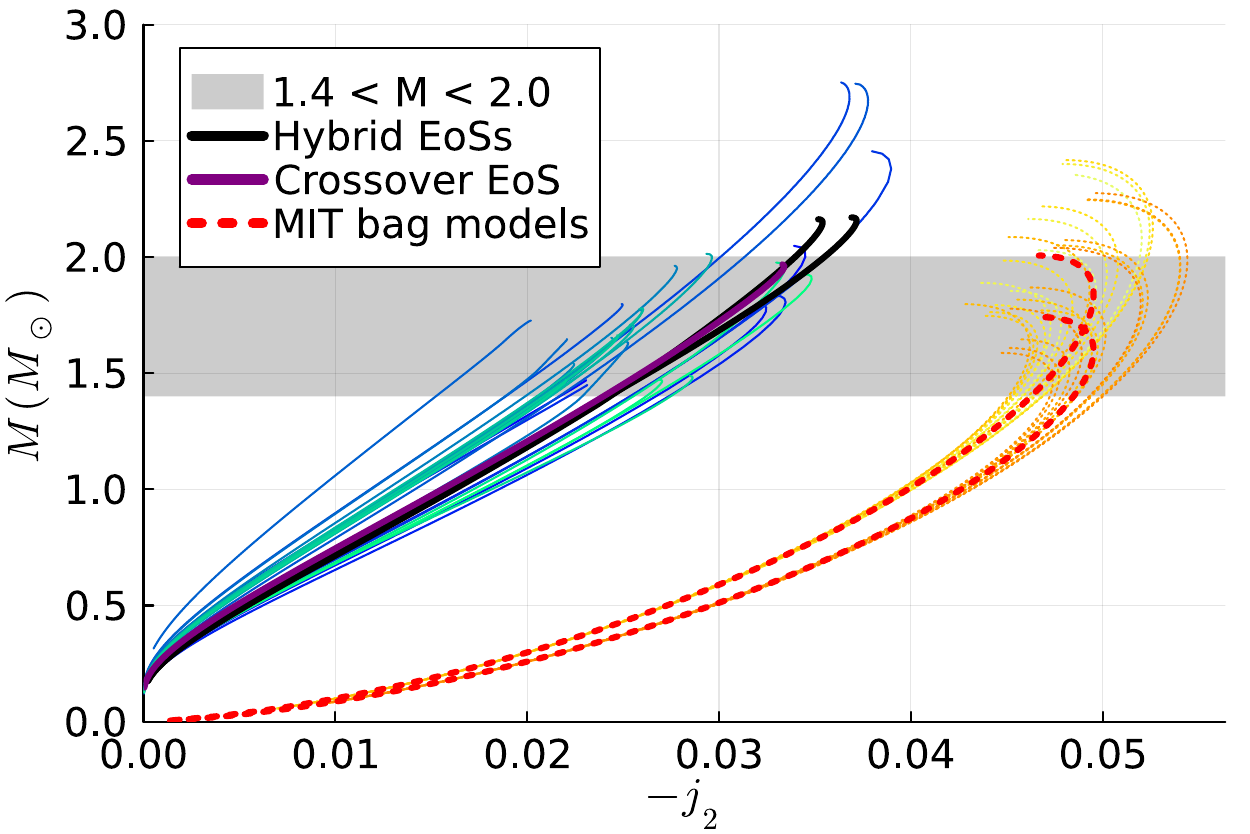}
    \caption{$M$-$\kmag$ curves for NSs (solid) and QSs (dotted). 
    The horizontal axis shows (irrotational) $-\kmag$. 
    }
    \label{fig:mkmag}
\end{figure}

As shown in Fig.~\ref{fig:mkmag},  the curves for QSs and NSs do not intersect in the $M$-$\kmag$ plane.  
Therefore, it is possible to distinguish massive QSs from NSs if $\kmag$ can be determined observationally.
For the experimental confirmation of the QS, we do not have to measure $\kmag$ precisely, but a lower limit of $|\kmag|$ would be sufficient.
Within the range of candidate NS EoSs, we find that $|\kmag|\lesssim 0.04$ for the high-mass range, $1.4\,M_\odot \lesssim M \lesssim 2.0\,M_\odot$.   Therefore, it is unlikely that the compact object with $|\kmag| \gtrsim 0.04$ is the NS\@. 
A value as large as $|\kmag| \approx 0.05$, determined with significant confidence, would then provide strong evidence for a QS.

\vspace{0.5em}

\paragraph{Discussions:}
So far, we point out that, if QSs and NSs have similar values of $\{M,\, R\}$ in the high-mass region ($M \gtrsim 1.4\,M_\odot$), their $\{M,\, \kel\}$ are also similar.
Therefore, even if the precision to measure $\kel$ in future observations improves significantly, it will still be a difficult challenge to make a conclusive statement of whether the observed object is the QS or the NS based on the $M$-$R$ or $M$-$\kel$ relation.
In contrast, once the information about $M$-$\kmag$ is available, the comparison between QSs and NSs exhibits clear separation.  
Thus, $\kmag$ is a promising discriminator for the QS candidate hunting.

From observational perspectives, according to Ref.~\cite{JimenezForteza:2018rwr}, $\kmag$ could be detectable with third-generation GW detectors~\cite{Punturo:2010zza, Hild:2010id}, but the proposed test relies on an approximate universal relation between $\kmag$ and $\kel$~\cite{Yagi:2013sva, Gagnon-Bischoff:2017tnz, Pani:2018inf}.
Such a procedure loses independent information about $\kmag$ in the waveform. 
Thus, the machinery argued in Ref.~\cite{JimenezForteza:2018rwr} is not of practical use to determine  $\kmag$ and $\kel$ as independent parameters.

The method to measure $\{M,\, \kel,\, \kmag\}$ independently has not been established yet, while independent measurement of $\{M,\, R,\, \kel\}$ would be far more difficult, which should involve observation of electromagnetic and GWs simultaneously from the common compact star.
The potential advantage in our proposal is that, unlike $\{M,\, R,\, \kel\}$, we do not have to require simultaneous and different measurements for $\{M,\, \kel,\, \kmag\}$, because the necessary information is encoded in a single GW observation.

How to exploit the $j_2$ separation in practice is still under
discussion.
One possible use is a hypothesis test applied after
$(M,\Lambda,\Sigma)$ is measured for a single object.
Assume that the object is an NS.
The NS Love-$C$ relation then converts the measured $\Lambda$ into a
compactness $C_{\mathrm{NS}}$, and the measured $\Sigma$ yields
$j_2^{\mathrm{NS}} = 48\,\Sigma\,C_{\mathrm{NS}}^5$.
The candidate NS EoSs give
$|j_2| \lesssim 0.04$ for $1.4\,M_\odot \lesssim M \lesssim 2.0\,M_\odot$
(Fig.~\ref{fig:mkmag}).  
If $|j_2^{\mathrm{NS}}| \gtrsim 0.04$, the NS
assumption is disfavored and the object is likely a QS.

Here, let us assess a prospect to differentiate between QSs and NSs using GW observations. 
In principle, we can constrain $\kmag$ from GW observations, if the waveform is identified at arbitrary precision.
In the post-Newtonian (PN) expansion of the gravitational waveform, $\kmag$ first contributes at 6PN order, whereas $\kel$ first contributes at 5PN order.
For waveform modeling, we parameterize the binary components by the masses $M_i$, the electric-type tidal deformabilities $\Lambda_i$, and the magnetic-type tidal deformabilities $\Sigma_i$.
To access $\kmag$, we must discriminate a QS--QS or QS--NS system from an NS--NS system in six-dimensional parameter space;  $(M_1,\, \Lambda_1,\, \Sigma_1)$ and $(M_2,\, \Lambda_2,\, \Sigma_2)$.  If the masses are first fixed, parameter space reduces to four dimensions, i.e., space spanned by $(\Lambda_1,\, \Sigma_1)$ and $(\Lambda_2,\, \Sigma_2)$.

Based on the method explicated in Ref.~\cite{Baird:2012cu}, 
we can estimate the minimally required signal-to-noise ratio (SNR) to distinguish between two different waveforms, $h_1(f)$ and $h_2(f)$.
With the power spectral density (PSD) denoted by $S_n(f)$ (which depends on the detector design), the noise-weighted inner product is defined by
\begin{equation}
    \langle h_1|h_2\rangle = 4\,\Re \int_{f_\mathrm{min}}^{f_{\mathrm{max}}} \frac{h_1(f)h_2^*(f)}{S_n(f)}\, df\,.
\end{equation}
We use the noise PSD, $S_n(f)$, from the ET-D design sensitivity curve~\cite{Hild:2010id}.
Using this inner product, the overlap, $F$, maximized with respect to the coalescence time $t_c$ and the coalescence phase $\phi_c$ is calculated as
\begin{equation}
    F(h_1, h_2) = \underset{t_c,\phi_c}{\max} \frac{\langle h_1|h_2\rangle}{\sqrt{\langle h_1|h_1\rangle}\sqrt{\langle h_2|h_2\rangle}}
\end{equation}
To distinguish these two waveforms in $\kappa$-dimensional parameter space [where $\kappa=6$ for $(M_i,\, \Lambda_i,\, \Sigma_i)$ and $\kappa=4$ for $(\Lambda_i,\, \Sigma_i)$], the following condition should be satisfied~\cite{Baird:2012cu}:
\begin{equation}
    2\rho^2 [1-F(h_1, h_2)] \geq  \chi^2_\kappa(1-p) \,,
\end{equation}
where $\rho$ is the SNR of our interest, and $\chi_\kappa^2(1-p)$ is the value of the $\chi$-squared distribution at cumulative probability $1-p$.
For a typical case, we compute $h_1=h_\mathrm{BNS}(f)$ from the Crossover EoS and $h_2=h_\mathrm{BQS}(f)$ from the MIT bag model EoS.
In both cases, the component masses are fixed to $M_1 = M_2 = 1.4\ M_\odot$, with which $(\Lambda_i,\, \Sigma_i)$ are uniquely derived from the EoSs.  We calculate the waveforms with TaylorF2 ($3.5$PN) plus multipolar tidal terms with ($\Lambda_i,\, \Sigma_i$); see Ref.~\cite{Henry:2020ski} for details.
The integration in Eq.~(5) uses $f_\mathrm{min} = 10\ \mathrm{Hz}$ and $f_\mathrm{max}$ set to the contact frequency ,$f_\mathrm{contact} = \sqrt{G(M_1+M_2)/(R_1+R_2)^3}/\pi$, of QS-QS\@.
Evaluating Eq.~(6) for these specific two waveforms yields $1-F = 8.910\times 10^{-4}$, from
which all the required SNR values below follow via Eq.~(7).
This is not a claim of measurability but a minimal, most-optimistic necessary check: whether the waveform difference between representative QS-QS and NS-NS systems is non-negligible in principle.
A full $(M, \Lambda, \Sigma)$ parameter-estimation analysis, including correlations and spins, is future work.
\begin{center}
    \begin{tabular}{c c c}
    \hline\hline
          $\kappa$ & ~~~$\rho$ $(68\%)$~~~ & $\rho$ $(90\%)$\\
    \hline
    $4$ & 51.3 & 66.1 \\
    $6$ & 62.7 & 77.3 \\
    \hline\hline
    \end{tabular}
\end{center}
These required $\rho$ values indicate that, to distinguish the binary QS merger from the binary NS merger using GW observations, one needs either a lower detector noise $S_n(f)$ or closer/louder events. 
For scale reference, the combined SNR of GW170817 was $32.4$ ($18.8$, $26.4$, and $2.0$ in LIGO–Hanford, LIGO–Livingston, and Virgo, respectively)~\cite{LIGOScientific:2017vwq}.
Our threshold is $51.3$ in the most optimistic case, which is about twice the GW170817 value. 
Given that this estimate is based on the ET-D design sensitivity curve, the present estimation of SNR should be viewed as a benchmark for third-generation detectors or later, rather than for the current detectors.
Alternatively, the $\Sigma$ effect can be quantified via the GW phase difference rather than SNR. 
\vspace{0.5em}

\paragraph{Summary:}
We found that $\kmag$ takes substantially different values for QSs and NSs even if their masses and radii are almost degenerate.  
Our findings should extend the frontier of the QS search toward higher mass regions and thus enhance the opportunity to identify the candidates with next-generation detectors.
We pointed out that the most notable advantage of considering $\kmag$ is that we do not have to make a simultaneous measurement, but a single GW signal conveys all information.
As a minimal, most-optimistic feasibility check, we estimated the SNR necessary for distinguishing the binary QS merger from the binary NS merger.
From our results, we can conclude that the idea of differentiating QS candidates among seemingly NS-like objects with $\kmag$ is promising.  
The potential of magnetic-type Love number would deserve further investigations along the lines of more realistic simulations in the future.

\section{Acknowledgments}
The authors thank
Toru~Kojo and
Shuhei~Minato
for useful discussions.
They are grateful to Hajime~Sotani
for a warm hospitality at Kochi University where a part of this work was completed.
This work was supported by the Japan Society for the Promotion of Science (JSPS) KAKENHI Grant Nos.\ 22H01216, 25K24464, 26K00698 (K.F.), FoPM, WINGS Program, The University of Tokyo (T.U.), and IIW, WINGS Program, The University of Tokyo (J.M.).
\bibliographystyle{apsrev4-2}

\bibliography{tidal}
\end{document}